\documentclass[showpacs,showkeys,preprintnumbers]{revtex4}%
\usepackage{amsfonts}
\usepackage{amsmath}
\usepackage{graphicx}
\usepackage{dcolumn}
\usepackage{bm}
\usepackage{amssymb}%
\begin{document}
\title{Hamilton-Jacobi Treatment of Superstring and Quantization of Fields with Constraints }
\author{Walaa I.\ Eshraim}
\affiliation{ Institute for Theoretical Physics, Goethe University,
Max-von-Laue-Str.\ 1, 60438 Frankfurt am Main, Germany, Institute for Theoretical Physics, Justus-Liebig University, Heinrich-Buff-Ring 16, 35392 Giessen, Germany.}

\begin{abstract}
The Hamilton-Jacobi formalism of constrained systems
 is used to study superstring. That obtained the equations of motion for a singular system
 as total differential equations in many variables. These
 equations of motion are in exact agreement with those equations
 obtained using Dirac's method. Moreover,
the Hamilton-Jacobi quantization of a constrained system is
 discussed. Quantization of the relativistic local free field with linear velocity of dimension D containing second-class constraints is studied.
 The set of Hamilton-Jacobi partial differential equations and the
 path integral of these theories are obtained by using the canonical
 path integral quantization. We figured out that the Hamilton-Jacobi path integral quantization of this system is in exact agreement with that given by using Senjanovic method. Furthermore, Hamilton-Jacobi path integral quantization of the scalar field coupled to two flavours of
fermions through Yukawa couplings is obtained directly as an integration over the canonical phase
space.
Hamilton-Jacobi quantization is applied to the constraint field systems with finite degrees of freedom by investigating the integrability conditions without using any gauge fixing condition.

\end{abstract}

\maketitle

\section{Introduction}
The generalized Hamiltonian dynamics describing
systems with constraints was initiated by Dirac \cite{Dirac0, Dirac1}, who established a formalism for treating constraint singular systems. The presence of constraints in
such theories requires care when applying Dirac's method,
especially when first-class constraints arise since the
first-class constraints are generators of gauge transformations
which lead to the gauge freedom. Dirac showed that the algebra of
Poisson brackets determines a division of constraints into two
classes: so-called first-class and second-class constraints. The
first-class constraints are those that have zero Poisson brackets
with all other constraints in the subspace of phase space in which
constraints hold;
constraints which are by definition second-class. Most physicists believe that this distinction is quite
important not only in classical theories but also in quantum
mechanics \cite{el1, el2}.\\
\indent As a first step in the present work, we intend to study a singular system with
Lagrangian describing superstring from the point of view of the
Hamilton-Jacobi formalism which has been developed by G\"{u}ler \cite{guler1, guler2} to investigate constrained systems. The equivalent Lagrangian
method \cite{guler3} is used to obtain the set of Hamilton-Jacobi partial
differential equations (HJPDE). The study
of such systems through Dirac's
generalized Hamiltonian formalism has already been extensively developed in literature \cite{el1, el2} to investigate theoretical models in contemporary elementary particle physics and will be used here for comparative purposes.\\
\indent Despite the success of Dirac's approach in studying
singular systems, which is demonstrated by the wide number of
physical systems to which this formalism has been applied, it is instructive to study singular systems through other
formalisms, since different procedures will provide different
views for the same problems, even for nonsingular systems. The
Hamilton-Jacobi approach that we study in this work, is applied
to some physical examples \cite{FG1, MG1, Bal1, WF0, WF1, WF2}. But it is still lacking to a better understanding
of this approach utility in the studying singular systems, and such understanding can only be achieved through its
application to other interesting physical systems. From our aims in this
work is to treat the superstring constraint system by the Hamilton-Jacobi approach and compare the results to those
obtained through Dirac's method.\\
\indent In the case of unconstrained systems, the Hamilton-Jacobi theory provides a bridge between classical and quantum mechanics.
The first study of the Hamilton-Jacobi equations for arbitrary
first-order actions was initiated by Santilli \cite{santilli}. The
quantization and construction of functional integral for theories
with first-class constraints in canonical gauge was given by
Faddeev and Popov \cite{Fadd, Pop}. Faddeev's method is generalized by Senjanovic \cite{Senj} to
the case of appearance of the second-class constraints in the theory.
Moreover, Fradkin \cite{Frad} considered quantization of bosonic theories
with first- and second-class constraints and the extension to
include in such gauges. Gitman and Tyutin \cite{el1} discussed the Hamiltonian formalism of gauge theories in an arbitrary gauge and the canonical quantization of singular theories.
In the Hamiltonian-Jacobi approach, the distinction
between first- and second-class constraints is not necessary. The
equations of motion are written as total differential equations in
many variables, which need to investigate the integrability conditions. In other words, the integrability conditions may lead
to new constraints. Moreover, it is shown that gauge fixing,
which is a basic procedure to study singular systems by
Dirac's method, is not necessary if the canonical method is used
\cite{guler3}. The path integral formulation based on the canonical method
is obtained in Refs. \cite{Bal2, MFH, WF3, WPath}.\\
\indent In Ref. \cite{WF4}, We have studied successfully the first-class constraints in canonical gauge by applying the Faddeev and Hamilton-Jacobi methods to obtain the path integral quantization of the scalar field coupled minimally to the vector potential. That led to the same results by the two methods which prove that the Hamilton-Jacobi method apple to quantized the first-class constraints. For more complement of confirmation the successful of Hamilton-Jacobi method, we quantize the relativistic local free field with a linear velocity of dimension D with the second-class constraints. The path integral quantization of this field is obtained by using the Senjanovic and Hamilton-Jacobi methods. We noticed that Faddeev \cite{WF4}, Popov and Senjanovic treatment need gauge-fixing conditions to obtain the path integral over the canonical variable, which is not always an easy task. However, if the Hamilton-Jacobi approach \cite{guler1, guler2} is used, the gauge fixing is not necessary to analyze singular systems \cite{MG1}. From the previous comparison, we figure out the ability and simplicity of using the Hamilton-Jacobi approach for studying the constraint systems. So, in the end, we get the path integral quantization of the scalar field coupled to two flavours of fermions through Yukawa couplings by using Hamilton-Jacobi quantization.\\
\indent This paper is organized as follows. In Sec. II we present the Hamilton-Jacobi approach. In Sec. III we present the path integral quantization methods which are Senjanovic's method and Hamilton-Jacobi quantization. In Sec. IV we treat the superstring constraint system by Dirac's method and Hamilton-Jacobi method. In Sec. V we quantize the relativistic local free field theory by using Senjanovic's method and Hamilton-Jacobi quantization. In Sec. VI we discuss the Hamilton-Jacobi quantization of the scalar field coupled to two flavours of fermions through Yukawa couplings. In Sec. VII we outline our conclusions.


\section{Hamilton-Jacobi Approach}

\qquad In this section, we approach the constrained systems by
Hamilton-Jacobi treatment, which solve the gauge-fixing problem naturally.\\
G\"{u}ler \cite{guler1, guler2} has developed a completely different method to
investigate singular systems. He started with the Hess matrix elements $A_{ik}$ of second derivatives of the Lagrangian $\mathcal{L}=\mathcal{L}(\varphi_i,\,\partial\varphi_i,\, \tau\,)$, $i=1,...,n$, which defined as 
\begin{equation}
A_{ik}= \frac{\delta^2\mathcal{L}(\varphi_i,\,\partial\varphi_i,\, \tau\,)}{\delta(\partial\varphi_i)\delta(\partial\varphi_k)}\,,\,\,\,\,\,\,\,\,i,k=1,2,...,n\,,\label{Hess}
\end{equation}
of rank $(n-r),\, r<n$, with dependent momenta $r$. The equivalent Lagrangian method \cite{guler3} is used to
obtain the set of Hamilton-Jacobi Partial Differential Equations
(HJPDE). The generalized momenta corresponding to generalized
coordinates $\varphi_i$ are defined as
\begin{align}
\pi_{a}&=
\frac{\overrightarrow{\delta}\mathcal{L}}{\delta(\partial_\mu\varphi_a)},\qquad
{a =1,2,\ldots,n-r}\,, \label{CM1}\\
\pi_{j}&=\frac{\overrightarrow{\delta}\mathcal{L}}{\delta(\partial_\mu\varphi_j)},\qquad
{j = n-r+1,\ldots,n}\,,\label{CM2}
\end{align}
where $\varphi_{i}$ are divided into two sets $\varphi_{a}$ and
$x_j$. Since the rank of Hess matrix is $(n-r)$, one may solve Eq.(\ref{CM1}) for $\partial_\mu \varphi_{a}$ as
\begin{equation}\label{Var1}
\partial_\mu\varphi_a=\partial_\mu\varphi_a(\varphi_i,\pi_a,\partial_\mu\varphi_j;\chi_\mu)\equiv
\omega_{a}\,,
\end{equation}
By substituting Eq. (\ref{Var1}) into Eq. (\ref{CM2}), we get
\begin{equation}\label{1.34}
\pi_{j}=
\frac{\overrightarrow{\delta}\mathcal{L}}{\delta(\partial_\mu\varphi_j)}\bigg|_{{\partial_\mu\varphi_a}
=\omega_a}\equiv -\mathcal{H}_{j}
(\varphi_{i},\partial_\mu\varphi_\nu,\pi_a;\chi_\mu)\,,
\end{equation}
which indicate to the fact that the generalized momenta
$\pi_{j}$ depend on $\pi_{a}$. That is a normal
result of the singular nature of the Lagrangian.
\\The canonical Hamiltonian $\mathcal{H}_{0}$ is given by the following definition
\begin{equation}
\mathcal{H}_{0}=-\mathcal{L}(\varphi_{i},\partial_\mu\varphi_\nu,\partial_\mu\varphi_{a}\equiv\omega_a,
\chi_\mu)+ \pi_{a}\omega_{a}+
\pi_{j}\partial_{\mu}\varphi_j\big|_{\pi_\nu=-\mathcal{H}_\nu}\,.\label{CH}
\end{equation}
The expression of the set of Hamilton-Jacobi Partial Differential Equations
(HJPDE) is 
\begin{equation}\label{HJP1}
\mathcal{H}'_0\bigg(\tau,~\varphi_{\nu},~\varphi_{a},~\pi_{i}=\frac{\overrightarrow{\delta}
S}{\delta \varphi_{i}} ,~\pi_{0}=\frac{\overrightarrow{\delta}
S}{\delta\chi_\mu}\bigg)=0,
\end{equation}
\begin{equation}\label{HJP2}
\mathcal{H}'_\mu
\bigg(\tau,~\varphi_{\nu},~\varphi_{a},~\pi_{i}=\frac{\overrightarrow{\delta}
S}{\delta \varphi_{i}} ,~\pi_{0}=\frac{\overrightarrow{\delta}
S}{\delta\chi_\mu}\bigg)=0,
\end{equation}
where $S$ being the action.\\
Eqs. (\ref{HJP1}) and (\ref{HJP2}) may be expressed in a compact form as
\begin{equation}\label{H}
\mathcal{H}'_\alpha
\bigg(\tau,~\varphi_{\nu},~\varphi_{a},~\pi_{i}=\frac{\overrightarrow{\delta}
S}{\delta \varphi_{i}} ,~\pi_{0}=\frac{\overrightarrow{\delta}
S}{\delta\chi_\alpha}\bigg)=0,
\end{equation}
$$\alpha = 0, n-r+1,\ldots,n\,.$$
where
\begin{align}\label{1.39}
\mathcal{H}'_0 = &\pi_{0}+\mathcal{H}_{0}=0\,, \\
\mathcal{H}'_\mu=& \pi_{j}+\mathcal{H}_{j} = 0\,.
\end{align}
Here $\mathcal{H}'_{0}$ can be interpreted as the generator of
time evolution while $\mathcal{H}'_{j}$ are the generators of
gauge transformation.
\\The fundamental equations of the equivalent
Lagrangian method are
\begin{align}\label{eqv}
\pi_{0}  &=\frac{\overrightarrow{\delta}
S}{\overrightarrow{\delta}\chi_\mu}\equiv
-\mathcal{H}_0(\varphi_i,\delta_\mu\varphi_\nu,\pi_a; \chi_\mu)\,,\\
 \pi_{a} &= \frac{\overrightarrow{\delta} S}{\delta
\varphi_{a}}\,, \,\,\,\,\,\,\,\,\,\pi_j = \frac{\overrightarrow{\delta}
S}{\delta \varphi_j}\equiv -\mathcal{H}_j\,,
\end{align}
with $ \varphi_{0}=\chi_\mu$. That gives the equations of motion as total
differential equations in many variables as 
\begin{align}
d\varphi_{r}&=\,\frac{\overrightarrow{\delta}
\mathcal{H}'_{\alpha}}{\delta \pi_{r}}\,
d\chi_{\alpha},\qquad \qquad\,\,\, r= 0,1,\ldots,n \, ,\label{EM1}\\
d\pi_{a}&=-\frac{\overrightarrow{\delta}
\mathcal{H}'_{\alpha}}{\delta \varphi_{a}}\,
d\chi_{\alpha},\qquad \qquad a=1,\ldots,n-r\,,\label{EM2}\\
d\pi_{\mu}&=-\frac{\overrightarrow{\delta}
\mathcal{H}'_{\alpha}}{\delta
\varphi_{\mu}}\,d\chi_{\alpha},\qquad \qquad \mu
=n-r+1,\ldots,n\,,\label{EM3}
\end{align}
\begin{equation}
dZ=\left(-\mathcal{H}_{\alpha}+\pi_{a}\frac{\overrightarrow{\delta}
\mathcal{H}'_{\alpha}}{\delta
\pi_{a}}\right)\,d\chi_{\alpha},\,\,\,\,\,\, \alpha = 0, n-r+1, \ldots,n\,,\label{EM4}
\end{equation}

where $Z=S(\chi_{\alpha},\varphi_{a})$. These equations are
integrable if and only if
\begin{align}
d\mathcal{H}'_0 &= 0\,,\label{IC1}\\
d\mathcal{H}'_\beta&= 0,\qquad ~~~~~~~~~\beta=1,2,\ldots,r\,.\label{IC2}
\end{align}
In the case of not satisfied the conditions (\ref{IC1}) and (\ref{IC2}) identically, one has to consider them as new constraints and then we examine again the variations of them. One is repeating this procedure until obtain
a set of conditions with all variations
vanish.\\
\indent The investigation of the integrability conditions \cite{Mu1, Mu2}
can be also done by using the operator method, where the linear
operators $X_\alpha$ corresponding to the set (\ref{EM1} - \ref{EM3}) are
defined as
\begin{equation}\label{O}
X_\alpha f (\chi_\beta, \varphi_a, \pi_a,z)=\frac{\delta f}{\delta
\chi_\alpha}+ \frac {\delta \mathcal{H}'_\alpha}{\delta \pi_a} \;
\frac{\delta f}{\delta \varphi_a}
 -\frac{\delta \mathcal{H}'_\alpha }{\delta \varphi_a} \; \frac{\delta f }{\delta
\pi_a}+\bigg(-\mathcal{H}_\alpha+\pi_a \frac{\delta
\mathcal{H}'_\alpha}{\delta \pi_a}\bigg)\frac{\delta f}{\delta
z}\,.
\end{equation}
The system is integrable, if the bracket relations
\begin{equation}\label{BR}
[X_\alpha,X_\beta] f = (X_\alpha X_\beta-X_\beta X_\alpha) f =
C_{\alpha \beta}^\gamma X_\gamma f;\quad \forall\,\, \alpha,\beta,\gamma=0,n-r+1,\ldots,n\,,
\end{equation}
are hold. If the relations (\ref{EM1} - \ref{EM3}) are not satisfied
identically, we add the bracket relations, which cannot be
expressed in this form as new operators. So the numbers of
independent operators are increased, and a new complete system can
be obtained. Then the new operators can be written in the Jacobi
form, and we fined the corresponding integrable system of the
total differential equations.


\section{Path Integral Quantization}
In this section, we briefly review the
Senjanovic's and Hamilton-Jacobi methods for studying the path
integral quantization of constrained systems.

\subsection{Senjanovic Method}
\quad We generalize Faddeeve's method \cite{Fadd} to the
case when second-class constraints are present. This
generalization is called Senjanovic's method.
\\Consider a mechanical system with $\alpha$ first-class
constraints $\phi_a$, $\beta$ second-class constraints $\theta_b$,
and the gauge conditions associated with the first-class
constraints $\chi_a$. Let the $\chi_a$ be chosen in such a way
that $\{\chi_a, \chi_b\}=0.$\\Then the expression for the $ S
$-matrix element is \cite{Senj}
\begin{equation}\label{Sej}
\left<Out\mid S\mid In\right>=\int exp
\left[i\int_{-\infty}^{\infty} (p_i \dot{q_i} - H_0)\, dt \right]
\prod_t \, d\mu (q(t), p(t))\,,
\end{equation}
and
\begin{equation}\label{3.4}
d\mu (q,p) = \left( \prod_{a=1}^\alpha \delta(\chi_a) \delta
(\phi_a) \right) det|| \{ \chi_a, \phi_a\}||\times
\prod_{b=1}^\beta \delta(\theta_b) \, det ||\{ \theta_a,
\theta_b\}||^\frac{1}{2} \prod_{i=1}^n dp_i \, dq^i\,.
\end{equation}
where $H_0$ is the Hamiltonian of the system and $d\mu (q,p)$ is
the measure of integration.

\subsection{Hamilton-Jacobi Quantization}

In Refs. \cite{Bal2, MFH, WF3, WF4, Mu1, Mu2, MG2}, the path integral
formulation of the constrained systems is studied. For computing the Hamilton-Jacobi path integral, one has to consider a singular Lagrangian as seen in section II. The canonical Hamiltonian $H_0$ defined in Eq. (\ref{CH}), and the set of
HJPDE is expressed in Eqs. (\ref{HJP1}) and (\ref{HJP2}). As we define
$p_{\beta}=\frac{\partial S[q_a;x_\alpha]}{\partial x_\beta}$ and
$p_a=\frac{\partial S[q_a;x_\alpha]}{\partial q_a}$ with $x_0=t$
and $S$ being the action. The total differential equations given
in (\ref{EM1} - \ref{EM4}) are integrable if (\ref{IC1}) and (\ref{IC2}) are hold \cite{MG2}.
If conditions  (\ref{IC1}) and (\ref{IC2}) are not satisfied identically, one
considers them as new constraints and again
consider their variations.\\
\indent Thus, repeating this procedure one may obtain a set of
constraints such that all variations vanish. Simultaneous
solutions of canonical equations with all these constraints
provide to obtain the set of canonical phase space coordinates
$(q_a,p_a)$ as functions of $t_\alpha$, besides the canonical
action integral is obtained in terms of the canonical coordinates.
$H'_\alpha$ can be interpreted as the infinitesimal generator of
canonical transformations given by parameters $t_\alpha$. In this
case path integral representation may be written as 
\begin{multline}\label{HJQ}
\left<Out\mid S\mid In\right> = \int \prod^{n-p}_{a=1}dq^a dp^a
exp \left\{\int_{t_\alpha}^{t'_\alpha} \left(-H_\alpha + p_\alpha
\frac{\partial H'_\alpha}{\partial p_\alpha}\right)\, dt_\alpha
\right\},\\\qquad a=1,\ldots,n-p,\qquad \alpha=0,n-p+1,\ldots,n\,.
\end{multline}
\indent In fact, this path integral is an integration over the
canonical phase-space coordinates $(q^a, p^a)$.

\section{Hamilton-Jacobi treatment of superstring}

\indent  In this section we treat supersting constraint system by Dirac's method and then apply Hamilton-Jacobi method.

\subsection{Dirac's formulation of superstring}
Consider a Lagrangian describes a superstring system
\begin{equation}\label{LSU}
L=-\frac{1}{2\pi}\,(\partial_{\alpha}X^{\mu}\partial^{\alpha}X_{\mu}-i\,{\overline{\psi}}\,^{\mu}\rho^{\alpha}\partial_{\alpha}\psi_{\mu})
-\jmath^{\alpha}A_{\alpha}-\frac{1}{4\pi}F^{\alpha\beta}F_{\alpha\beta}\,,
\end{equation}
Where $A^{\alpha}$ is a world sheet potential analogous to the
electromagnetic potential. The world sheet current density
\begin{equation}\label{J1}
\jmath_{\alpha}=\frac{1}{2\pi}\,q\,{\overline{\psi}}\,^{\mu}\rho^{\alpha}\psi_{\mu}\,,
\end{equation}
acts as a source for the gauge field $(A^{\alpha})$, and the electromagnetic tensor is defined as $F^{\alpha\beta}=\partial^{\alpha}A^{\beta}-\partial^{\beta}A^{\alpha}$.\\
\indent The Lagrangian (\ref{LSU}) is singular, since the rank of the
Hess matrix (\ref{Hess}) is four. The generalized momenta (\ref{CM1}) and (\ref{CM2}) can
be written as
\begin{equation}\label{ML1}
p_{\mu}=\frac{\partial
L}{\partial{\dot{X}}^{\mu}}=-\frac{1}{\pi}\,\partial^{0}X_{\mu}\,,
\end{equation}
\begin{equation}\label{ML2}
p^{\mu}_{\overline\psi} =\frac{\partial
L}{\partial{\dot{\overline\psi\,^{\mu}}}}=0=-H_{\overline{\psi}}\,,
\end{equation}
\begin{equation}\label{ML3}
p^{\mu}_{\psi}=\frac{\partial
L}{\partial{\dot{\psi}}_{\mu}}=\frac{i}{2\pi}{{\overline\psi\,^\mu}}{\rho^0}=
-H_{\psi}\,,
\end{equation}
\begin{equation}\label{ML4}
\pi^{i} = \frac{\partial
L}{\partial{\dot{A}}_i}=\frac{1}{\pi}F^{i0}\,,
\end{equation}
\begin{equation}\label{ML5}
\pi^0 = \frac{\partial L}{\partial{\dot{A}}_0}=0=-H_1\,.
\end{equation}
Equations (\ref{ML1}) and (\ref{ML4}), respectively leads us to express the
velocities $\dot{X}^\mu$ and $\dot{A}_i$ as
\begin{equation}\label{24}
\dot{X}^\mu=-\pi\,p^{\mu}\,,
\end{equation}
\begin{equation}\label{25}
\dot{A}_i=-\pi\,\pi_{i}+\partial_{i}A_{0}\,.
\end{equation}
The Hamiltonian density is given by
\begin{equation}\label{26}
 H_0 =-{\frac{\pi}{2}}(p^\mu\,p_\mu+\pi^i\,\pi_i)+\pi^i\,\partial_{i}A_{0}+
 {\frac{1}{2\pi}}[\partial_{i}X^{\mu}\,\partial^{i}X_{\mu}-
 i{\overline{\psi}}\,^{\mu}\rho^i\,\partial_{i}\psi_{\mu}+q\overline{\psi}\,^{\mu}\rho^{\alpha}\psi_{\mu}A_{\alpha}+
\frac{1}{2}F^{ij}\,F_{ij}]\,.
\end{equation}
The total Hamiltonian density is constructed as
\begin{align}
 H_T =&-{\frac{\pi}{2}}(p^\mu\,p_\mu+\pi^i\,\pi_i)+\pi^i\,\partial_{i}A_{0}+
 {\frac{1}{2\pi}}[\partial_{i}X^{\mu}\,\partial^{i}X_{\mu}-
 i\overline{\psi}\,^{\mu}(\rho^i\,\partial_{i}\nonumber\\ &+iq\rho^{\alpha}A_{\alpha})\psi_{\mu}+
\frac{1}{2}F^{ij}\,F_{ij}]+\lambda_{\overline{\psi}}p^{\mu}_{\overline{\psi}}+\lambda_{\psi}(p^{\mu}_\psi-\frac{i}{2\pi}\overline{\psi}^{\mu}\rho^0)+\lambda_{1}\pi_{0}\,,\label{THL1}
\end{align}
where $\lambda_{\overline{\psi}}$, $\lambda_{\psi}$ and
$\lambda_{1}$ are Lagrange multipliers to be determined. From the
consistency conditions, the time derivative of the primary
constraints should be zero, that is
\begin{equation}\label{td1}
\dot{H'}_{\overline{\psi}}=\{H'_{\overline{\psi}}\,,H_T\}=\frac{1}{2\pi}(i\rho^{i}\partial_{i}-q\rho^{\alpha}\,A_{\alpha})\psi_{\mu}+\frac{i}{2\pi}\rho^0\,\lambda_{\psi}\approx0\,,
\end{equation}
\begin{equation}\label{td2}
\dot{H'}_{\psi}=\{H'_{\psi}\,,H_T\}=-\frac{1}{2\pi}\overline{\psi}\,^{\mu}(i\overleftarrow{\partial_{i}}\rho^{i}+q\rho^{\alpha}\,A_{\alpha})-\frac{i}{2\pi}\lambda_{\overline{\psi}}\rho^0\,\approx0\,,
\end{equation}
\begin{equation}\label{td3}
\dot{H'}_{1}=\{H'_{1},H_T\}=\partial_{i}\pi^i-\frac{1}{2\pi}\,q\overline{\psi}\,^{\mu}\rho^0\psi_{\mu}\approx0\,.
\end{equation}
Relations (\ref{td1}) and (\ref{td2}) fix the multipliers $\lambda_{\psi}$ and
$\lambda_{\overline{\psi}}$ respectively as
\begin{equation}\label{Lame1}
\lambda_{\psi}=-(
\rho^{0}\,\rho^{i}\,\partial_{i}+iq\,\rho^{0}\,\rho^{\alpha}\,A_{\alpha})\,\psi_{\mu}\,,
\end{equation}
\begin{equation}\label{Lame2}
\lambda_{\overline{\psi}}=-\overline{\psi}\,^{\mu}(\overleftarrow{\partial_{i}}\,\rho^i\,\rho^0-iq\,\rho^{\alpha}\,\rho^0\,A_\alpha)\,.
\end{equation}
Eq. (\ref{td3}) lead to the secondary constraints
\begin{equation}\label{33}
H''_{1}=\partial_{i}\pi^i-\frac{1}{2\pi}\,q\overline{\psi}\,^{\mu}\rho^0\psi_{\mu}\approx0\,.
\end{equation}
There are no tertiary constraints, since
\begin{equation}\label{34}
\dot{H''}_{1}=\{H''_{1},H_T\}=0\,.
\end{equation}
By taking suitable linear combinations of constraints, one has to
find the first-class, that is
\begin{equation}\label{35}
\Phi_{1}=H'_{1}= \pi_0\,,
\end{equation}
whereas the constraints
\begin{equation}\label{36}
\Phi_{2}= H'_{\overline\psi}=p_{\overline\psi}^\mu\,,
\end{equation}
\begin{equation}\label{37}
\Phi_{3}=
H'_{\psi}=p_{\psi}^\mu-\frac{i}{2\pi}\,\overline{\psi}\,^{\mu}\rho^0\,,
\end{equation}
\begin{equation}\label{38}
\Phi_{4}=H''_{1}=\partial_{i}\pi^i-\frac{1}{2\pi}\,q\overline{\psi}\,^{\mu}\rho^0\psi_{\mu}\,,
\end{equation}
are second-class.\\
The equations of motion read as
\begin{equation}\label{EML1}
\dot{X}^\mu=\{X^\mu,H_T\}=-\pi\,p^{\mu}\,,
\end{equation}

\begin{equation}\label{EML2}
\dot{\overline{\psi}}\,^\mu=\{{\overline{\psi}}\,^\mu,H_T\}=\lambda_{\overline{\psi}}\,,
\end{equation}

\begin{equation}\label{EML3}
\dot{\psi}\,^\mu=\{\psi_\mu,H_T\}=\lambda_{\psi}\,,
\end{equation}

\begin{equation}\label{EML4}
\dot{A}^i=\{A^i,H_T\}=-(\pi\,\pi^{i}-\partial_{i}A_{0})\,,
\end{equation}

\begin{equation}\label{EML5}
\dot{A}^0=\{A^0,H_T\}=\lambda_1\,,
\end{equation}

\begin{equation}\label{EML6}
{\dot{p}_\mu}=\{p_\mu,H_T\}=\frac{1}{\pi}\,\partial^{i}\partial_{i}X^\mu\,,
\end{equation}

\begin{equation}\label{EML7}
{\dot{p}}_{\overline{\psi}}^\mu=\{p_{\overline{\psi}}^\mu\,,H_T\}=\frac{1}{2\pi}\,[(i\rho^i\,\partial_i-q\rho^\alpha\,A_\alpha)\psi_\mu+i\lambda_\psi\,\rho^0]\,,
\end{equation}

\begin{equation}\label{EML8}
{\dot{p}}_{\psi}^\mu=\{p_{\psi}^\mu\,,H_T\}=-\frac{i}{2\pi}\,{\overline{\psi}}\,^\mu(\overleftarrow{\partial_{i}}\rho^i+iq\,\rho^\alpha\,A_\alpha)\,,
\end{equation}

\begin{equation}\label{EML9}
{\dot{\pi}}^i=\{\pi^i,H_T\}=\frac{1}{2\pi}\,(2\partial^lF_{il}-q\,{\overline{\psi}}\,^\mu\,\rho^i\,\psi_\mu)\,,
\end{equation}

\begin{equation}\label{EML10}
{\dot{\pi}}^0=\{\pi^0,H_T\}=\partial_i\pi^i\,-\frac{1}{2\pi}\,q\,{\overline{\psi}}\,^{\mu}\,\rho^{0}\,\psi_{\mu}\,.
\end{equation}

Substituting from Eq. (\ref{Lame2}) into Eq. (\ref{EML2}), we get
\begin{equation}\label{49}
i{\overline{\psi}}\,^{\mu}\rho^\alpha(\overleftarrow{\partial_\alpha}-iqA_{\alpha})=0\,,
\end{equation}
and from Eq. (\ref{Lame1}) into Eqs. (\ref{EML3}) and (\ref{EML7}), we have
\begin{equation}\label{50}
i(\partial_\alpha+iqA_{\alpha})\rho^{\alpha}\,\psi_{\mu}=0\,,
\end{equation}

\begin{equation}\label{51}
{\dot{p}}_{\overline \psi}^\mu=0\,.
\end{equation}
\indent We will contact oureselves with a partial gauge fixing by
introducing gauge constraints for the first class primary
constraints only, just to fix the multiplier $\lambda_1$ in Eq.
(\ref{THL1}). Since $\pi^0$ is vanishing weakly, a gauge choice near at
hand would be
\begin{equation}\label{52}
\Phi'_1=A_0=0\,.
\end{equation}
But for this forbids dynamics at all, since the requirement
$\dot{A}_0=0$ implies $\lambda_1=0$.\\
 In the following section the same system
will be discussed using Hamilton-Jacobi approach.

\subsection{Hamilton-Jacobi formulation of superstring}
The set of Hamilton-Jacobi Partial Differential Equations (HJPDE)
(\ref{HJP1}) read as
\begin{equation}\label{53}
H'_0 =p_0+H_0=0\,,
\end{equation}
\begin{equation}\label{54}
H'_{\overline\psi}=p_{\overline
\psi}^\mu+H_{\overline\psi}=p_{\overline \psi}^\mu=0\,,
\end{equation}
\begin{equation}\label{55}
H'_{\psi}=p_{\psi}^\mu+H_{\psi}=p_{\psi}^\mu-\frac{i}{2\pi}\>{\overline{\psi}}\,^\mu{\rho^0}=0\,,
\end{equation}
\begin{equation}\label{56}
H'_1 =\pi_0+H_1=\pi_0=0\,.
\end{equation}

\indent The equations of motion are obtained as total differential
equations follows:
\begin{equation}
 dX^\mu=\frac{\partial H'_{0}}{\partial
p_{\mu}}\>dt+\frac{\partial H'_{\overline{\psi}}}{\partial
p_{\mu}}\>d\overline{\psi}\,^\mu+\frac{\partial
H'_{\psi}}{\partial p_{\mu}}\>d\psi_\mu+\frac{\partial
H'_1}{\partial p_{\mu}}\>dA^0
=-\pi\,p^{\mu}\,dt\,,\label{Tde1}
\end{equation}

\begin{equation}
 dA^{i}=\frac{\partial H'_{0}}{\partial
\pi_{i}}\>dt+\frac{\partial H'_{\overline{\psi}}}{\partial
\pi_{i}}\>d\overline{\psi}\,^\mu+\frac{\partial
H'_{\psi}}{\partial \pi_{i}}\>d\psi_\mu+\frac{\partial
H'_1}{\partial \pi_{i}}\>dA^0
=-(\pi\,\pi^{i}-\partial_{i}A_{0})\,dt\,,\label{Tde2}
\end{equation}

\begin{equation}
dp_{\mu}=-\frac{\partial H'_{0}}{\partial
X_{\mu}}\>dt-\frac{\partial H'_{\overline{\psi}}}{\partial
X_{\mu}}\>d\overline{\psi}\,^\mu-\frac{\partial
H'_{\psi}}{\partial X_{\mu}}\>d\psi_\mu-\frac{\partial
H'_1}{\partial
X_{\mu}}\>dA^0=\frac{1}{\pi}\,\partial^{i}\partial_{i}X^\mu\,dt\,,\label{Tde3}
\end{equation}

\begin{equation}
dp^{\mu}_{\overline{\psi}}=-\frac{\partial
H'_{0}}{\partial \overline{\psi}\,^\mu}\>dt-\frac{\partial
H'_{\overline{\psi}}}{\partial \overline{\psi}\,^\mu
}\>d\overline{\psi}\,^\mu-\frac{\partial H'_{\psi}}{\partial
\overline{\psi}\,^\mu}\>d\psi_\mu-\frac{\partial H'_1}{\partial
\overline{\psi}\,^\mu}\>dA^0
=\frac{1}{2\pi}\,(i\rho^i\,\partial_i-q\rho^\alpha\,A_\alpha)\psi_\mu\,dt+\frac{i}{2\pi}\,\rho^0
d\psi_{\mu}\,, \label{Tde4}
\end{equation}

\begin{equation}
dp_{\psi}^\mu=-\frac{\partial
H'_{0}}{\partial \psi_{\mu}}\>dt-\frac{\partial
H'_{\overline{\psi}}}{\partial \psi_{\mu}
}\>d\overline{\psi}\,^\mu-\frac{\partial H'_{\psi}}{\partial
\psi_{\mu}}\>d\psi_\mu-\frac{\partial H'_1}{\partial
\psi_{\mu}}\>dA^0=-\frac{1}{2\pi}(i\,\partial_{i}\,{\overline{\psi}}\,^\mu\,\rho^i+q\,{\overline{\psi}}\,^\mu\,\rho^\alpha\,A_\alpha)\,dt \,,\label{Tde5}
\end{equation}

\begin{equation}
 d\pi^i=-\frac{\partial H'_{0}}{\partial
A_i}\>dt-\frac{\partial H'_{\overline{\psi}}}{\partial A_i
}\>d\overline{\psi}\,^\mu-\frac{\partial H'_{\psi}}{\partial A_i
}\>d\psi_\mu-\frac{\partial H'_1}{\partial A_i}\>dA^0
=\frac{1}{2\pi}\,(2\partial^lF_{il}-q\,{\overline{\psi}}\,^\mu\,\rho^i\,\psi_\mu)\,dt\,,\label{Tde6}
\end{equation}

\begin{equation}
 d\pi^0=-\frac{\partial H'_{0}}{\partial
A_0}\>dt-\frac{\partial H'_{\overline{\psi}}}{\partial A_0
}\>d\overline{\psi}\,^\mu-\frac{\partial H'_{\psi}}{\partial A_0
}\>d\psi_\mu-\frac{\partial H'_1}{\partial A_0}\>dA^0
=(\partial_i\pi^i\,-\frac{1}{2\pi}\,q\,{\overline{\psi}}\,^{\mu}\,\rho^{0}\,\psi_{\mu})dt\,.\label{Tde7}
\end{equation}
\indent The integrability conditions imply that the variation of
the constraints $H'_{\overline{\psi}}$, $H'_{\psi}$ and $H'_{1}$
should be identically zero; that is
\begin{equation}\label{dhl1}
dH'_{\overline\psi}=dp_{\overline \psi}^\mu=0\,,
\end{equation}

\begin{equation}\label{dhl2}
dH'_{\psi}=dp_{\psi}^\mu-\frac{i}{2\pi}\>d{\overline{\psi}}\,^\mu{\rho^0}=0\,,
\end{equation}

\begin{equation}\label{dhl3}
dH'_1 =d\pi_0=0\,.
\end{equation}
The vanishing of total differential of $H'_{1}$ leads to a new
constraints
\begin{equation}\label{67}
H''_{1}=\partial_{i}\pi^i-\frac{1}{2\pi}\,q\overline{\psi}\,^{\mu}\rho^0\psi_{\mu}\,.
\end{equation}
When we taking a gain the total differential of $H''_{1}$, we
notice that it vanishes identically,
\begin{equation}\label{68}
dH''_{1}=0\,.
\end{equation}
From Eqs. (\ref{Tde1}) and (\ref{Tde2}), respectively we obtain
\begin{equation}\label{69}
\dot{X}^\mu=-\pi\,p^{\mu}\,,
\end{equation}
and
\begin{equation}\label{70}
\dot{A}^i=-(\pi\,\pi^{i}-\partial_{i}A_{0})\,.
\end{equation}
Substituting from Eqs. (\ref{Tde4}) and (\ref{Tde5}) into Eqs. (\ref{dhl1}) and (\ref{dhl2})
respectively, we get
\begin{equation}\label{EO1}
i(\partial_\alpha+iqA_{\alpha})\rho^{\alpha}\,\psi_{\mu}=0\,,
\end{equation}

\begin{equation}\label{72}
i{\overline{\psi}}\,^{\mu}\rho^\alpha(\overleftarrow{\partial_\alpha}-iqA_{\alpha})=0\,.
\end{equation}
Also from Eqs. (\ref{Tde3}) and (\ref{Tde5} - \ref{Tde7}), we get the following equations of
motion:
\begin{equation}\label{73}
{\dot{p}_\mu}=\frac{1}{\pi}\,\partial^{i}\partial_{i}X^\mu\,,
\end{equation}

\begin{equation}\label{74}
{\dot{p}}_{\psi}^\mu=-\frac{i}{2\pi}\,{\overline{\psi}}\,^\mu(\overleftarrow{\partial_{i}}\rho^i+iq\,\rho^\alpha\,A_\alpha)\,,
\end{equation}

\begin{equation}\label{75}
{\dot{\pi}}^i=\frac{1}{2\pi}\,(2\partial^lF_{il}-q\,{\overline{\psi}}\,^\mu\,\rho^i\,\psi_\mu)\,,
\end{equation}

\begin{equation}\label{76}
{\dot{\pi}}^0=\partial_i\pi^i\,-\frac{1}{2\pi}\,q\,{\overline{\psi}}\,^{\mu}\,\rho^{0}\,\psi_{\mu}\,.
\end{equation}
Substituting from Eq.(\ref{EO1}) into Eq.(\ref{Tde4}), we have
\begin{equation}\label{77}
{\dot{p}}_{\overline{\psi}}^{\mu}= 0\,.
\end{equation}
As a comparison between the above two methods, we get that the
Hamilton-Jacobi method and Dirac's method give the same equations
of motion.


\section{Path Integral Quantization of The Relativistic Local Free Field Theory}

As an example of a singular system described by a first order
action, namely a system whose lagrange function is linear in the
velocities. However, the associated constraints are all
second-class. Let us consider the relativistic local free field
theory of spin $\frac{1}{2}$ in a Minkowski spacetime of dimension
D. As usual, spacetime coordinates are denoted as
$x^{\mu},y^{\mu}(\mu=0,1,\ldots,D-1)$ and space components are
labelled by  $i,j = 1,2,\ldots,D-1$. The Minkowski matrix
${\eta^{{\mu}{\nu}}}$ is chosen with a signature with mostly minus
signs, and we also set $\hbar=c=1$. The system is described by the
first order action
\begin{equation}\label{17}
S[\psi]={\int{d^D}x\>l(\psi,{\partial_{\mu}\psi})}.
\end{equation}
with the local lagrangian density function
\begin{equation}\label{L2}
l(\psi,{\partial_{\mu}\psi})=i{\frac{\lambda+1}{2}}\>{\overline{\psi}}{\gamma^\mu}{\partial_{\mu}\psi}
+i{\frac{\lambda-1}{2}}\>{\partial_{\mu}{\overline{\psi}}}{\gamma^\mu}{\psi}-m{\overline{\psi}}{\psi}\,.
\end{equation}
Here $\lambda$ is a parameter, the matrices $\gamma^\mu$ define
the Dirac algebra in D-dimensional Minkowski spacetime
\begin{equation}\label{19}
\{{\gamma^\mu},{\gamma^\nu}\}=2\>{\eta^{\mu\nu}},\>\>\>{\gamma^{\mu\dag}}={\gamma^0}{\gamma^\mu}{\gamma^0}\,,
\end{equation}
and ${\psi_{\alpha}(x)}(\alpha=1,2,\dots,2^{[{D}/{2}]})$ are
Grassmann even degrees of freedom defining a Dirac spinor, with
\begin{equation}\label{20}
{\overline{\psi}}(x)={\psi^\dag}(x)\>{\gamma^0}\,.
\end{equation}
For simplicity, the fields $\psi(x)$ are assumed to fall off
sufficiently rapidly at infinity for all practical purposes.\\
The Lagrangian (\ref{L2}) is singular, since the rank of the Hess matrix
(\ref{Hess}) is zero.\\
Let us first discuss the system using Hamilton-Jacobi approach. In
this approach the canonical momenta (\ref{CM1}) and (\ref{CM2}) take the forms

\begin{equation}\label{21}
p=\frac{\partial L}{\partial{\partial_0}\psi}=
i{\frac{\lambda+1}{2}}\>{\overline{\psi}}{\gamma^0}= -H\,,
\end{equation}
and
\begin{equation}\label{22}
\overline{p} =\frac{\partial
L}{\partial{\partial_0}\overline{\psi}}=
i{\frac{\lambda-1}{2}}\>{\gamma^0}{\psi}= -\overline{H}\,.
\end{equation}
where we must call attention to the necessity of being careful
with the spinor indexes. Considering, as usual $\psi$ as a column
vector and $\overline{\psi}$ as a row vector implies that $p$ will
be a row vector while $\overline{p}$ will be a column vector.\\
The usual Hamiltonian $H_0$ is given as
\begin{equation}\label{23}
H_0 = -L+{{\partial_0}\psi}\>p_{\mu}+
{{\partial_0}\overline{\psi}}\>p_{\overline{\mu}}\>\bigg|_{{p_{\mu}=-H_{\mu}},\>
{p_{\overline{\mu}= -H_{\overline{\mu}}}}}\,,
\end{equation}
or,
\begin{equation}\label{24}
H_0 =
-i{\frac{\lambda+1}{2}}\>{\overline{\psi}}{\gamma^a}{\partial_a
\psi}-i{\frac{\lambda-1}{2}}\>{\partial_a
{\overline{\psi}}}{\gamma^a}{\psi}+m{\overline{\psi}}{\psi}\,,\,\,\,\,\,\,\,\,\, a=1,2,3\,.
\end{equation}

The set of Hamilton-Jacobi partial differential equation (HJPDE)
are
\begin{equation}\label{HJPL21}
H'_0 =p_0+H_0=p_0
-i{\frac{\lambda+1}{2}}\>{\overline{\psi}}{\gamma^a}{\partial_a
\psi}-i{\frac{\lambda-1}{2}}\>{\partial_a
{\overline{\psi}}}{\gamma^a}{\psi}+m{\overline{\psi}}{\psi}\,,
\end{equation}
\begin{equation}\label{HJPL22}
H'= p+H= p-i{\frac{\lambda+1}{2}}\>{\overline{\psi}}{\gamma^0}=0\,,
\end{equation}
\begin{equation}\label{HJPL23}
\overline{H'}= \overline{p}+\overline{H}=
\overline{p}-i{\frac{\lambda-1}{2}}\>{\gamma^0}\psi=0\,.
\end{equation}
Therefor, the total differential equations for the characteristic
(\ref{EM1} - \ref{EM3}) are:
\begin{equation}\label{TDL21}
d\psi=d\psi\,,
\end{equation}
\begin{equation}\label{TDL22}
d{\overline\psi}=d{\overline\psi}\,,
\end{equation}
\begin{equation}\label{TDL23}
d{p}=\bigg(i{\frac{\lambda-1}{2}}\>{\partial_a
{\overline{\psi}}}{\gamma^a}-m{\overline{\psi}}\bigg)d\tau+i{\frac{\lambda-1}{2}}\>{\gamma^0}
d{\overline{\psi}}\,,
\end{equation}
\begin{equation}\label{TDL24}
d{\overline{p}}=\bigg(i{\frac{\lambda+1}{2}}\>{\partial_a
{\psi}}{\gamma^a}-m{\psi}\bigg)d\tau+i{\frac{\lambda+1}{2}}\>{\gamma^0}
d{\psi}\,.
\end{equation}
To check wether the set of equations (\ref{TDL21} - \ref{TDL24}) is integrable or
not, we have to consider the total variation of the constraints.
In fact
\begin{equation}\label{32}
dH'=dp-i{\frac{\lambda+1}{2}}\>{\gamma^0}d{\overline{\psi}}=0\,,
\end{equation}
\begin{equation}\label{33}
d\overline{H'}=d\overline{p}-i{\frac{\lambda-1}{2}}\>{\gamma^0}d\psi=0\,.
\end{equation}
The constraints (\ref{HJPL22})  and (\ref{HJPL23}), lead us to obtain
\begin{equation}\label{34}
d\overline{\psi}=i(i\partial_a\overline{\psi}\gamma^a+m\overline{\psi})\gamma^0dt\,,
\end{equation}
and
\begin{equation}\label{35}
d\psi=i\gamma^0(i\gamma^a\partial_a\psi-m\psi)dt\,.
\end{equation}
Then, we conclude that the set of equations  (\ref{TDL21} - \ref{TDL24}) is
integrable.\\
\indent Making use of Eq. (\ref{EM4}) and  Eqs. (\ref{HJPL21} - \ref{HJPL23}), we can write the
canonical action integral as
\begin{equation}\label{36}
Z=\int d^4x
\bigg(i{\frac{\lambda+1}{2}}\>{\overline{\psi}}{\gamma^\mu}{\partial_\mu
\psi}+i{\frac{\lambda-1}{2}}\>{\partial_\mu
{\overline{\psi}}}{\gamma^\mu}{\psi}-m{\overline{\psi}}{\psi}\bigg)\,.
\end{equation}
Now the S-matrix element is given by
\begin{equation}\label{IntL2}
\left<\psi,\overline{\psi},\iota;\psi',\overline{\psi}',\iota\right>
= \int d\psi d\overline{\psi} exp\bigg[i\bigg\{\int
d^{4}x\bigg(i{\frac{\lambda+1}{2}}\>{\overline{\psi}}{\gamma^\mu}{\partial_\mu
\psi}+i{\frac{\lambda-1}{2}}\>{\partial_\mu
{\overline{\psi}}}{\gamma^\mu}{\psi}-m{\overline{\psi}}{\psi}\bigg)\bigg\}
\bigg]\,.
\end{equation}
\indent Now we will apply the Senjanovic method to the previous
example.\\
The total Hamiltonian is given as
\begin{equation}\label{38}
H_T=H_0+\nu H'+ \overline{\nu} \overline{H'}\,,
\end{equation}
or
\begin{equation}\label{39}
H_T=
-i{\frac{\lambda+1}{2}}\>{\overline{\psi}}{\gamma^a}{\partial_a
\psi}-i{\frac{\lambda-1}{2}}\>{\partial_a
{\overline{\psi}}}{\gamma^a}{\psi}+m{\overline{\psi}}{\psi}
+\nu(p-i{\frac{\lambda+1}{2}}\>{\overline{\psi}}{\gamma^0})
+\overline{\nu}(\overline{p}-i{\frac{\lambda-1}{2}}\>{\gamma^0}\psi)\,,
\end{equation}
where $\nu$ and $\overline{\nu}$ are Lagrange multipliers to be
determined. From the consistency conditions, the time derivative
of the primary constraints should be zero, that is
\begin{equation}\label{dc1}
\dot{H'}=\{H',H_T\}=-i\partial_a\overline{\psi}\,{\gamma^a}-m{\overline{\psi}}-i\,\overline{\nu}\,{\gamma^0}\,\approx0\,,
\end{equation}
\begin{equation}\label{dc2}
\dot{\overline{H}'}=\{\overline{H}',H_T\}=i\,{\gamma^a}\,\partial_a\psi-m\,\psi+i\,{\gamma^0}\nu\,\approx0\,.
\end{equation}
Eqs. (\ref{dc1}) and (\ref{dc2}) fix the multipliers $\overline{\nu}$ and
$\nu$, respectively as
\begin{equation}\label{42}
\overline{\nu}=-\overline{\psi}(\overleftarrow{\partial_a}\gamma^a-im)\gamma^0\,,
\end{equation}
and
\begin{equation}\label{43}
\nu=-\gamma^0(\gamma^a\,\partial_a+im)\>{\psi}\,.
\end{equation}
There are no secondary constraints. By taking suitable linear
combinations of constraints, one has to find the maximal number of
second class only, there are
\begin{equation}\label{44}
\Phi_1=H'= p -i{\frac{\lambda+1}{2}}\>{\overline{\psi}}{\gamma^0}\,,
\end{equation}
and
\begin{equation}\label{45}
\Phi_2=\overline{H}'=
\overline{p}-i{\frac{\lambda-1}{2}}\>{\gamma^0}\psi\,.
\end{equation}
The total Hamiltonian is vanishing weakly. It can completely be
written in terms of second-class constraints as
\begin{equation}\label{46}
H_T=
-i{\frac{\lambda+1}{2}}\>{\overline{\psi}}{\gamma^a}{\partial_a
\psi}-i{\frac{\lambda-1}{2}}\>{\partial_a
{\overline{\psi}}}{\gamma^a}{\psi}+m{\overline{\psi}}{\psi}
+\nu\,\Phi_1 +\overline{\nu}\Phi_2\,.
\end{equation}
The equations of motion are read as
\begin{equation}\label{47}
\dot{\psi}=\{\psi,H_T\}=\nu\,,
\end{equation}
\begin{equation}\label{48}
\dot{\overline{\psi}}=\{\overline{\psi},H_T\}=\overline{\nu}\,,
\end{equation}
\begin{equation}\label{49}
\dot{p}=\{p,H_T\}=-i{\partial_a{\overline{\psi}}}{\gamma^a}-m{\overline{\psi}}
+i{\overline{\nu}}\>{\frac{\lambda-1}{2}}\>{\gamma^0}\,,
\end{equation}
and
\begin{equation}\label{50}
\dot{\overline{p}}=\{\overline{p},H_T\}=i{\gamma^a}{\partial_a{\psi}}-m{\psi}
+i\nu\>\frac{\lambda+1}{2}\>{\gamma^0}\,.
\end{equation}
\indent To obtain the path integral quantization, taking into our
consideration that we have two constraints (primary constraint),
which are second-class constraints, then we make use the
Senjanovic method Eq. (\ref{Sej}) one obtains
\begin{multline}\label{51}
\left<Out|S|In\right> = \int exp\bigg[i \int^{+\infty}_{-\infty}
\bigg(i{\frac{\lambda+1}{2}}\>{\overline{\psi}}{\gamma^\mu}{\partial_\mu
\psi}+i{\frac{\lambda-1}{2}}\>{\partial_\mu{\overline{\psi}}}{\gamma^\mu}{\psi}-m{\overline{\psi}}{\psi}\bigg)\bigg]dt\,D\psi\,Dp
D\overline{\psi}\,\>D\overline{p}\,\> det(\gamma^0
I)\\\times\>\delta(p-i{\frac{\lambda+1}{2}}\>{\overline{\psi}}{\gamma^0})\>\delta(\overline{p}-i{\frac{\lambda-1}{2}}\gamma^0\psi)\,.
\end{multline}
After integrating over $p$ and $\overline{p}$ one can arrive at
the result which has seen in Eq. (\ref{IntL2}).


\section{Hamilton-Jacobi quantization of the scalar
field coupled to two fLavours of fermions through yukawa couplings} 
We consider one loop order the self-energy for the
scalar field $\varphi$ with a mass $m$, coupled to two flavours of
fermions with masses $m{_1}$ and $m{_2}$, coupled through Yukawa
couplings described by the lagrangian
\begin{equation}\label{L3}
L={\frac{1}{2}}(\partial_{\mu}\varphi)^2-{\frac{1}{2}}{m^2}{\varphi^2}-{\frac{1}{6}}\lambda
{\varphi^3}+\sum_{i}{{\overline{\psi}}_{(i)}}(i{\gamma^\mu}\partial_\mu-m_i){\psi_{(i)}}
-g\varphi({{\overline{\psi}}_{(1)}}{\psi_{(2)}}+{{\overline{\psi}}_{(2)}}{\psi_{(1)}}),
\qquad{\mu =0,1,2,3}\,,
\end{equation}
where $\lambda$ is parameter and $g$ constant, $\varphi$,
$\psi_{(i)}$, and ${\overline\psi}_{(i)}$ are odd ones. We are
adopting the Minkowski metric
$\eta_{\mu\nu}=diag(+1,-1,-1,-1)$.\\
\indent The Lagrangian function (\ref{L3}) is singular, since the rank
of the Hess matrix (\ref{Hess}) is one. The generalized momenta (\ref{CM1}, \ref{CM2}) are
\begin{equation}\label{gm1}
p_{\varphi} =\frac{\partial
L}{\partial\dot{\varphi}}={\partial^0}\varphi\,,
\end{equation}
\begin{equation}\label{18}
p_{(i)} =\frac{\partial L}{\partial{\dot{\psi}}_{(i)}}=
i{{\overline{\psi}}_{(i)}}{\gamma^0}= -H_{(i)},\qquad {i =1,2}\,,
\end{equation}
\begin{equation}\label{19}
\overline{p}_{(i)}=\frac{\partial
L}{\partial{\dot{\overline\psi}}_{(i)}}=0=-\overline{H}_{(i)}\,.
\end{equation}
Where we must call attention to the necessity of being careful
with the spinor indexes. Considering, as usual $\psi_{(i)}$ as a
column vector and ${\overline\psi}_{(i)}$ as a row vector implies
that $p_{(i)}$ will be a row vector while ${\overline p}_{(i)}$
will be a column vector. \\
 Since the rank of the Hess matrix is one, one may
solve (\ref{gm1}) for ${\partial^0}\varphi$ as
\begin{equation}\label{20}
{\partial^0}\varphi=p_{\varphi}\equiv\omega\,.
\end{equation}
The usual Hamiltonian $H_0$ is given as
\begin{equation}\label{21}
H_0 = -L+\omega
p_{\varphi}+{{\partial_0}\psi_{(i)}}\>p_{(i)}\>{\bigg|_{{p_{(i)}=-H_{(i)}}}}+{
{{\partial_0}{\overline{\psi}_{(i)}}}\>{\overline
p}_{(i)}}\>\bigg|_ {{\overline p}_{(i)}= -{\overline H}_{(i)}}\,,
\end{equation}
or
\begin{equation}\label{22}
H_0 =
{\frac{1}{2}}({p^2}_{\varphi}-{\partial_{a}\varphi}{\partial^a}\varphi)+{\frac{1}{2}}{m^2}{\varphi^2}+{\frac{1}{6}}\lambda
{\varphi^3}-{{\overline{\psi}}_{(i)}}(i{\gamma^a}\partial_{a}-m_i){\psi_{(i)}}+g\varphi({{\overline{\psi}}_{(1)}}{\psi_{(2)}}+{{\overline{\psi}}_{(2)}}{\psi_{(1)}})\,,\qquad {a=1,2,3}\,.
\end{equation}
The set of Hamilton-Jacobi Partial Differential Equations
(\ref{HJP1}) and (\ref{HJP2}) read as
\begin{align}\label{HJPL31}
H'_0 &=p_0+H_0\nonumber\\
&=p_0+{\frac{1}{2}}({p^2}_{\varphi}-{\partial_{a}\varphi}{\partial^a}\varphi)+{\frac{1}{2}}{m^2}{\varphi^2}+{\frac{1}{6}}\lambda
{\varphi^3}-{{\overline{\psi}}_{(i)}}(i{\gamma^a}\partial_{a}-m_i){\psi_{(i)}}
+g\varphi({{\overline{\psi}}_{(1)}}{\psi_{(2)}}+{{\overline{\psi}}_{(2)}}{\psi_{(1)}})\,,
\end{align}
\begin{equation}\label{HJP32}
H'_{(i)}=p_{(i)}+H_{(i)}=p_{(i)}-i\>{{\overline{\psi}}_{(i)}}\,{\gamma^0}=0\,,
\end{equation}

\begin{equation}\label{HJP33}
\overline H'_{(i)}=\overline p_{(i)}+\overline H_{(i)}=\overline
p_{(i)}=0\,.
\end{equation}

Therefor, the total differential equations for the characteristic
(\ref{EM1} - \ref{EM3}) are:
\begin{equation}\label{deL31}
d\varphi=p_{\varphi}d\tau\,,
\end{equation}

\begin{equation}\label{deL32}
d{\psi_{(i)}}=d{\psi_{(i)}}\,,
\end{equation}

\begin{equation}\label{deL33}
d{{\overline\psi}_{(i)}}=d{{\overline\psi}_{(i)}}\,,
\end{equation}

\begin{equation}\label{deL34}
d{{p}_{\varphi}}=\bigg[{m^2}\varphi+{\frac{1}{2}}\lambda{\varphi^2}
+g({{\overline{\psi}}_{(1)}}{\psi_{(2)}}+{{\overline{\psi}}_{(2)}}{\psi_{(1)}})\bigg]d\tau\,,
\end{equation}

\begin{equation}\label{deL35}
d{{p}_{(1)}}=\bigg[{\overline\psi}_{(1)}(i\overleftarrow{\partial_a}\gamma^a
+m_{1})+g\,\varphi\,{\overline\psi}_{(2)}\bigg]d\tau\,,
\end{equation}

\begin{equation}\label{deL36}
d{{p}_{(2)}}=\bigg[{\overline\psi}_{(2)}(i\overleftarrow{\partial_a}\gamma^a
+m_{2})+g\,\varphi\,{\overline\psi}_{(1)}\bigg]d\tau\,,
\end{equation}

\begin{equation}\label{deL37}
d{\overline
p_{(1)}}=\bigg[-(i{\gamma^a}{\partial_a}-m_{1}){\psi_{(1)}}+g\varphi{\psi_{(2)}}\bigg]d\tau
-i{\gamma^0}d{\psi_{(1)}}\,,
\end{equation}
and
\begin{equation}\label{deL38}
d{\overline
p_{(2)}}=\bigg[-(i{\gamma^a}{\partial_a}-m_{2}){\psi_{(2)}}+g\varphi{\psi_{(1)}}\bigg]d\tau
-i{\gamma^0}d{\psi_{(2)}}\,.
\end{equation}
\\
\indent To check whether the set of equations (\ref{deL31} - \ref{deL38}) is
integrable or not, we have to consider the total variations of the
constraints. In fact\\
\begin{equation}\label{34}
dH'_{(i)}=dp_{(i)}-i\>d{{\overline{\psi}}_{(i)}}\,{\gamma^0}=0\,,
\end{equation}

\begin{equation}\label{35}
d{\overline H'_{(i)}}=d\overline p_{(i)}=0\,.
\end{equation}
The constraints (\ref{HJP32}) and (\ref{HJP33}), lead us to obtain
$d\overline{\psi}_{(i)}$ and $d\psi_{(i)}$ in terms of $dt$

\begin{equation}\label{36}
d{\overline\psi}_{(1)}i\gamma^{0}=[{\overline\psi}_{(1)}(i\overleftarrow{\partial_a}\gamma^a
+m_{1})+g\varphi\,{\overline\psi}_{(2)}]dt\,,
\end{equation}

\begin{equation}\label{37}
d{\overline\psi}_{(2)}i\gamma^{0}=[{\overline\psi}_{(2)}(i\overleftarrow{\partial_a}\gamma^a
+m_{2})+g\varphi\,{\overline\psi}_{(1)}]dt\,,
\end{equation}

\begin{equation}\label{38}
i{\gamma^0}d\psi_{(1)}=[-(i\gamma^a{\partial_a}-m_{1}){\psi_{(1)}}+g\,\varphi\>{\psi_{(2)}}]d\,,
\end{equation}
and
\begin{equation}\label{39}
i{\gamma^0}d\psi_{(2)}=[-(i\gamma^a{\partial_a}-m_{2}){\psi_{(2)}}+g\,\varphi\>{\psi_{(1)}}]dt\,.
\end{equation}
 We obtain that the set of equations (\ref{deL31} - \ref{deL38})  is integrable. Making use of
(\ref{EM4}), and  (\ref{HJPL31} - \ref{HJP33}), we can write the canonical action integral as
\begin{equation}\label{40}
\quad Z=\int
d^{4}x[\frac{1}{2}({p^2}_{\varphi}+{\partial_{a}\varphi}{\partial^a}\varphi)-{\frac{1}{2}}{m^2}{\varphi^2}-{\frac{1}{6}}\lambda
{\varphi^3}+{{\overline{\psi}}_{(i)}}(i{\gamma^\mu}\partial_{\mu}-m_i){\psi_{(i)}}
-g\varphi({{\overline{\psi}}_{(1)}}{\psi_{(2)}}+{{\overline{\psi}}_{(2)}}{\psi_{(1)}})]\,,
\end{equation}
Now the path integral representation (\ref{HJQ}) is given by
\begin{align}\label{44}
\left<out|S|In\right> = \int\prod_{i}^2\, d\varphi\,
dp_{\varphi}\,d\psi_{(i)}\,d\overline{\psi}_{(i)}\>exp\,\bigg\{i\bigg[\int
d^{4}x
\frac{1}{2}({p^2}_{\varphi}+{\partial_{a}\varphi}{\partial^a}\varphi)-{\frac{1}{2}}{m^2}{\varphi^2}&-{\frac{1}{6}}\lambda
{\varphi^3}+{{\overline{\psi}}_{(i)}}(i{\gamma^\mu}\partial_{\mu}-m_i){\psi_{(i)}}\nonumber\\
&-g\varphi({{\overline{\psi}}_{(1)}}{\psi_{(2)}}+{{\overline{\psi}}_{(2)}}{\psi_{(1)}})\bigg]\bigg\}\,.
\end{align}

\section {Conclusion}

\qquad In this paper, we have investigated three different constrained systems. Two of them are studied by using Dirac's Hamiltonian formalism and Hamilton-Jacobi approach. The third one quantized by Hamilton-Jacobi quantization.\\
\indent We have treated constrained system of the Lagrangian describing superstring and have obtained the equations of motion of this system by Dirac's and Hamilton-Jacobi method. In the Dirac method the total Hamiltonian composed by adding
the constraints multiplied by Lagrange multipliers to the
canonical Hamiltonian. In order to drive the equations of motion,
one needs to redefine these unknown
multipliers in an arbitrary way. However, in the Hamilton-Jacobi formalism, there is no
need to introduce Lagrange multipliers to the canonical
Hamiltonian. Both the consistency conditions and integrability
conditions lead to the same constraints. In the Hamilton-Jacobi
formulation, the equations of motion are obtained directly by
using HJPDES as total differential equations. \\
\indent Path integral quantization of the relativistic local free
field theory is obtained by using the Senjanovic method and the Hamilton-Jacobi path integral
formulation. Both methods give the same results. However, in the Hamilton-Jacobi path integral formulation, since the integrability conditions $dH'$ and
$d\overline{H}'$ are satisfied, so this system is integrable, and
hence the path integral is obtained directly as an integration
over the canonical phase-space coordinates
$(\psi,\overline{\psi})$. In the usual formulation, one has to
integrate over the extended phase-space
$(p,\psi,\overline{p},\overline{\psi})$ and one can get rid of the
redundant variables $(p,\overline{p})$ by using delta function
$\delta(p - i\frac{\lambda+1}{2}\overline{\psi}\gamma^0)$ and
$\delta(\overline{p} - i\frac{\lambda-1}{2}\gamma^0\psi)$. 
\indent Furthermore, the scalar field coupled to two flavours of fermions through
Yukawa couplings are quantized as a constrained system by using
Hamilton-Jacobi quantization. That is no need to introduce
Lagrange multipliers to the canonical Hamiltonian, then the
Hamilton-Jacobi is simpler and more economical.\\
\indent As a conclusion, the Hamilton-Jacobi approach is always in
exact agreement with Dirac's method. Both the consistency
conditions and integrability conditions lead to the same
constraints. The singular system with second-class constraints is quantized by Hamilton-Jacobi quantization successfully. The Hamilton-Jacobi path integral
quantization is simpler and more economical. In Hamilton-Jacobi treatment, there is no need to
distinguish between first-class and second-class constraints, and there is
no need to introduce Lagrange multipliers; all that is needed is
the set of Hamilton-Jacobi partial differential equations and the
equations of motion. If the system is integrable then one can
construct the canonical phase space.
 In hamilton-Jacobi quantization, the gauge fixing is not necessary to obtain
the path integral formulation for field theories if the canonical
formulation is used. Since this system is integrable,
the path integral is obtained as an integration over the canonical
phase-space coordinates.

\end{document}